\documentclass[twocolumn,showpacs,preprintnumbers,amsmath,amssymb,prb]{revtex4-1}

\usepackage[active]{srcltx}
\usepackage{graphicx}% Include figure files
\usepackage{dcolumn}% Align table columns on decimal point
\usepackage{bm}% bold math

\usepackage{hyperref}
\usepackage{color}
\usepackage[a4paper]{anysize}

\definecolor{darkred}{rgb}{0.5,0.0,0.0}
\definecolor{darkgreen}{rgb}{0.0,0.5,0.0}
\definecolor{darkblue}{rgb}{0.0,0.0,0.5}
\hypersetup{colorlinks=true,
citecolor=darkred,
linkcolor=darkgreen,
urlcolor=darkblue}

\newcommand{\R}{{\bm{R}}}

\newcommand{\A}{{\bf A}}
\newcommand{\E}{{\bf E}}
\newcommand{\e}{{\bm{e}}}
\renewcommand{\a}{{\bf a}}
\renewcommand{\j}{{\bf j}}
\renewcommand{\k}{{\bm{k}}}

\newcommand{\J}{{\bf J}}
\newcommand{\q}{{\bm{q}}}
\newcommand{\p}{{\bm{p}}}
\renewcommand{\r}{{\bm{r}}}

\renewcommand{\v}{{\bm{v}}}
\newcommand{\I}{{\rm i}}

\newcommand{\re}{\ensuremath{\mathrm{Re}\,}}
\newcommand{\im}{\ensuremath{\mathrm{Im}\,}}

\def\gsim{\lower.35em\hbox{$\stackrel{\textstyle>}{\textstyle\sim}$}}
\def\lsim{\lower.35em\hbox{$\stackrel{\textstyle<}{\textstyle\sim}$}}
\begin{document}

\title{Universal absorption of two-dimensional systems}

\author{T. Stauber$^{1,2}$, D. Noriega-P\'erez$^{2,3}$, and J. Schliemann$^4$}
\affiliation{$^1$Departamento de Teor\'{\i}a y Simulaci\'on de Materiales, Instituto de Ciencia de Materiales de Madrid, CSIC, 28049 Cantoblanco, Spain\\
$^2$Departamento de F\'{\i}sica de la Materia Condensada, INC and IFIMAC, Universidad Aut\'onoma de Madrid, E-28049 Madrid, Spain\\
$^3$Departamento de F\'{\i}sica, Universidad de Oviedo, E-33007 Oviedo, Spain\\
$^4$Institute for Theoretical Physics, 
University of Regensburg, D-93040 Regensburg, Germany}

\begin{abstract}
We discuss the optical conductivity of several non-interacting two-dimensional (2D) semiconducting systems focusing on gapped Dirac and Schr\"odinger fermions as well as on a system mixing these two types. Close to the band-gap, we can define a universal optical conductivity quantum of $\sigma_0=\frac{1}{16}\frac{e^2}{\hbar}$ for the pure systems. The effective optical conductivity then depends on the degeneracy factors $g_s$ (spin) and $g_v$ (valley) and on the curvature around the band-gap $\nu$, i.e., it generally reads $\sigma=g_sg_v\nu\sigma_0$. For a system composed of both types of carriers, the optical conductivity becomes non-universal. 
\end{abstract}


%\noindent {\it keywords}: optical properties of graphene, 2D plasmons

\maketitle

%%%%%%%%%%%%%%%%%%%%%%%%%%%%%%%%%%%%%%%%%%%%%%%%%%%%%%%%%%%%%
%  SECTION INTRODUCTION
%%%%%%%%%%%%%%%%%%%%%%%%%%%%%%%%%%%%%%%%%%%%%%%%%%%%%%%%%%%%%
\section{Introduction}
Suspended graphene absorbs 2.3\% of the incoming energy flux over a broad frequency region ranging from the far-infrared to the visible regime of the spectrum.\cite{Mak08,Nair08} The numerical value is obtained from the  universal constants $\pi\alpha$ where $\alpha\approx1/137$ denotes the fine-structure constant which is related to the universal optical conductivity of graphene, $\sigma=\frac{e^2}{4\hbar}$.\cite{Ludwig94,Ando02,Gusynin06,FalkovskyPer07,Peres08} The universality is due to the cancellation of the Fermi velocity that appears in the density-of-states as well as in the band-overlap. This cancellation is exact within the Dirac model, but also approximately holds in the visible regime where trigonal warping effects need to be taken into account.\cite{Stauber08} And even vertex corrections due to electron-electron interactions hardly change this universal behavior.\cite{Mishchenko08,Sheehy09,Teber14}

Recent absorption experiments on InAs-monolayers show an absorption of $\mathcal{A}=\pi\alpha\mathcal{F}$ with the local field-correction $\mathcal{F}=\frac{4}{(1+n_s)^2}$ due to the substrate with refractive index $n_s$.\cite{Fang13}  This translates into an effective optical conductivity of $\sigma(\omega\approx\omega_\Delta)=\frac{e^2}{4\hbar}$ for transitions close to the frequency that corresponds to the band-gap, $\omega_\Delta$. There thus seems to be a universal absorption and optical conductivity in two-dimensional systems, respectively, independent of whether they are composed of chiral and gapless Dirac or gapped parabolic Schr\"odinger electrons.  

In this work, we shall investigate this intriguing universality in more detail and our results can be summarized as follows. One can define a minimal universal optical conductivity $\sigma_0=\frac{e^2}{16\hbar}$ giving rise to an absorption quantum $\mathcal{A}_0=\frac{\pi}{4}\alpha$ that should be observable, e.g., in 3D topological insulators.\cite{Hasan10,Qi11} The general optical conductivity and absorption is then given by $\sigma=g_sg_v\nu\sigma_0$ and $\mathcal{A}=g_sg_v\nu\mathcal{A}_0$, respectively, where $g_s$ and $g_v$ denote the spin and valley degeneracy and $\nu$ defines the curvature  around the band gap, $\epsilon_{v,c}\sim |k|^\nu$. 

The optical conductivity per channel of a gapped system consisting of parabolic Schr\"odinger Fermions is thus $\sigma_{channel}=2\sigma_0$ and twice as large as the one of a gapless system like graphene with $\sigma_{channel}=\sigma_0$. But for gapped Dirac Fermions, one also obtains $\sigma(\omega\approx\omega_\Delta)=2\sigma_0$, so that for a gapped system the optical conductivity per channel is $\sigma_{channel}=2\sigma_0$ independent of the chiral nature of its carriers. This equivalence is a necessary condition since for a large gap as, e.g., present in MoS$_2$, the Hamiltonian can either be written in terms of Pauli-matrices\cite{Rostami13} or in terms of parabolic Schr\"odinger fermions\cite{Berghauser14} used for typical semiconductors. 

The paper is organised as follows. In Sec. II, we will first generalize the formulas for the conductivity for chiral (Dirac) fermions with arbitrary curvature $\nu$. In Sec. III, we then derive the conductivity for Schr\"odinger Fermions. In Sec. IV, we will treat the hybrid system of Dirac and Schr\"odinger electrons which can be experimentally obtained in the case of transition metal dichalcogenides (e.g., MoS$_2$) or mercury telluride quantum wells (Te(Cd)Hg). We close with conclusions and an outlook.

\section{Conductivity of chiral Dirac electrons}
We discuss the optical conductivity per channel using the Kubo formula\cite{Fetter03}
\begin{align}
\label{Kubo}
\sigma^{ij}(\omega)=&-\frac{\I e^2}{(\omega+\I0) A}\sum_{m\neq n}\frac{n_F(\epsilon_m)-n_F(\epsilon_n)}{\hbar\omega-\epsilon_m+\epsilon_n+\I0}\\\notag
&\times\langle m|v^i|n\rangle\langle n|v^ j|m\rangle+\sigma^{ij}_{dia}\;, 
\end{align}
where $n,m$ label the eigenstates of the corresponding Hamiltonian and $A$ denotes the area of the system. The conductivity also contains the 
diamagnetic contribution $\sigma^{ij}_{dia}\sim \delta_{ij}$  which will be 
discuss in more detail in Sec. \ref{FullModelDia}. As we shall see there,
Dirac fermions do not contribute to the diamagnetic current which is entirely 
due to Schr\"odinger particles.
The velocity operator is given by
\begin{align}
\label{velocity}
\v=\frac{\I}{\hbar}[H,\r]=\frac{\partial H}{\partial \p}\;. 
\end{align}

For $T=0$, the real part of the longitudinal optical conductivity can then be written in the following form:
\begin{align}
\re\sigma^{ii}(\omega)=D\delta(\omega)+\re\sigma_{reg}^{ii}(\omega)
\end{align}
where $D$ denotes the Drude weight\footnote{We include the prefactor $\pi$ in the definition of the Drude weight} and the regular part is given by
\begin{align}
\re&\sigma_{reg}^{ii}(\omega)=\frac{16\pi\sigma_0}{\omega A}\sum_{m\neq n}
\langle m|v^i|n\rangle\langle n|v^i|m\rangle\\\notag
&\times\left(n_F(\epsilon_m)-n_F(\epsilon_n)\right)\delta(\omega-(\epsilon_m-\epsilon_n)/\hbar)
\end{align}
where we introduced the optical conductance quantum
\begin{align}
\sigma_0=\frac{1}{16}\frac{e^2}{\hbar} \;.
\end{align}
In the following, we will only discuss the longitudinal conductivity and will thus drop the super-indices $\sigma^{ii}\to\sigma$. 
\subsection{Hamiltonian}
Let us investigate a model Hamiltonian of chiral massive electrons
\begin{equation}
H= \gamma  \left( \begin{array}{cc}
\Delta/2 & (k_x -  \I k_y)^n  \\
(k_x +\I k_y)^n  & -\Delta/2 
\end{array} \right) \;.
\end{equation}
For $\gamma=\hbar v_F$ ($n=1$)and $\gamma=(\hbar v_F)^2/t_\perp$ ($n=2$), this is the effective low energy Hamiltonian for single and bilayer graphene with $\Delta=0$, respectively.\cite{McCann06} The eigenenergies are given by $\epsilon_\k^\pm=\pm\gamma\sqrt{k^{2n}+(\Delta/2)^2}$. For massive chiral electrons ($\Delta\neq0$), we thus have $\epsilon_\k^\pm\sim \pm k^{2n}$ ($\nu=2n$); for massless chiral electrons ($\Delta=0$), we have $\epsilon_\k^\pm\sim \pm k^{n}$ ($\nu=n$).

The spinor eigenvectors are given by 
\begin{align}
\label{eigenvectors}
|\k,+\rangle&= \left( \begin{array}{c}
\cos\vartheta/2\\
\sin\vartheta/2e^{\I n\phi}  
\end{array} \right)\;,\\
|\k,-\rangle&= \left( \begin{array}{c}
\sin\vartheta/2\\
-\cos\vartheta/2e^{\I n\phi}  
\end{array} \right)\;,
\end{align}
with $\cos\vartheta=\frac{\Delta}{\sqrt{4k^{2n}+\Delta^2}}$ and $\phi$ the angle between $\k$ and the $x$-axis.

The response function of the above Hamiltonian has been discussed extensively in the literature for the special cases $n=1,2,3$.\cite{Ando02,Gusynin06,Nilsson06,Koshino09} Still, we are unaware of any publication that emphasizes on the general case $n$ with respect to the universal absorption quantum. The following formulas are thus generalizations of what has already been presented, previously.

\subsection{Drude weight}
Let us first discuss the intraband contribution to the optical conductivity. This is most directly done within the density-density response function since no potential contribution due to the diamagnetic term needs to be taken into account. In the local approximation ($\q\to0$), there is no chiral band-overlap\footnote{This also holds true for the general model of Eq. (\ref{HybridModel}).} and the density-density correlation function per channel can be approximated for general isotropic dispersion as
\begin{align}
\chi_{\rho\rho}=\frac{1}{(2\pi)^2}\int d^2k\frac{n_F(\epsilon_\k^s)-n_F(\epsilon_{\k+\q}^s)}{\hbar\omega+\epsilon_\k^s-\epsilon_{\k+\q}^s}\;,
\end{align}
where a summation over the band-index $s=\pm$ is implied.
In the limit $\q\rightarrow0$, this becomes
\begin{equation}
\label{rhorho}
\chi_{\rho\rho}=\frac{1}{(2\pi)^2}\int d^2k\left(-\frac{\partial n_F(\epsilon_\k^s)}{\partial \epsilon_\k^s}\right)\left(\frac{\nabla \epsilon_\k^s\cdot\q}{\hbar\omega}\right)^2.
\end{equation}
With the Fermi frequency $\omega_\mu=2\mu/\hbar$ and gap frequency $\omega_\Delta=\gamma\Delta/\hbar$, the Drude weight defined by $D=\pi e^2\lim_{\omega\to0}\frac{\omega^2}{q^2}\chi_{\rho\rho}$ is then given at $T=0$ by
\begin{equation}
\label{DrudeWeight}
D=2n\sigma_0\omega_\mu\left(1-\left(\frac{\omega_\Delta}{\omega_\mu}\right)^2\right)\;.
\end{equation}
 
\subsection{Interband transitions}
For interband transitions, we need to evaluate the matrix elements involving the velocity operator. After integration over the angle, one obtains
\begin{align}
\frac{1}{2\pi}\int_0^{2\pi}|\langle \k,+|v^i|\k,-\rangle|^2=\left(\frac{n\gamma k^{n-1}}{\hbar}\right)^2\frac{1+\cos^2\vartheta}{2}\;.
\end{align}

This yields the following real part of the conductivity
\begin{align}
\re\sigma_{reg}(\omega) =n\sigma_0\left(1+\left(\frac{\omega_\Delta}{\omega}\right)^2\right) 
 \theta(\omega-\omega_{max})\;,
\end{align}
with $\omega_{max}=\max(\omega_\mu, \omega_\Delta)$. 

The imaginary part is modified by the same factor. This yields the following expression:
\begin{align}
\im\sigma(\omega)=\frac{n\sigma_0}{\pi}\left(1+\left(\frac{\omega_\Delta}{\omega}\right)^2\right)\ln\frac{\omega-\omega_{max}}{\omega+\omega_{max}}\;.
\end{align}
 
\subsection{Full conductivity}
We can now present the general formula of the longitudinal conductivity for the above model, including the degeneracy factors for the spin and valley degrees of freedom. This yields
\begin{widetext}
\begin{align}
\sigma(\omega)&=ng_sg_v\sigma_0\Big[2\omega_\mu\left(1-\left(\frac{\omega_\Delta}{\omega_\mu}\right)^2\right)\left(\delta(\omega)+\frac{\I}{\omega}\right)+\left(1+\left(\frac{\omega_\Delta}{\omega}\right)^2\right)\left(\theta(\omega-\omega_{max})+\frac{\I}{\pi}\ln\frac{\omega-\omega_{max}}{\omega+\omega_{max}}\right)\Big]\;.
\end{align}
\end{widetext}

The influence of finite temperature can be included\cite{FalkovskyPer07,Peres08,Stauber08} and by broadening the delta-function, one can also treat damping effects in a phenomenological way. With $\omega_T=2k_BT/\hbar$, we obtain the following expression:
\begin{widetext}
\begin{align}
\label{FiniteTemp}
\sigma(\omega)&=2ng_sg_v \sigma_0\Big[\frac{\I\omega_T}{\omega+\I\gamma_\tau}\frac{1}{\pi}\ln\left((e^{(-\omega_\mu+\omega_\Delta)/\omega_T}+1)(e^{(\omega_\mu+\omega_\Delta)/\omega_T}+1)\right)
+\frac{1}{4}\left(1+\left(\frac{\omega_\Delta}{\omega}\right)^2\right)\\\notag
&\times\left(\tanh\frac{\omega+\omega_{max}}{2\omega_T}+\tanh\frac{\omega-\omega_{max}}{2\omega_T}+\frac{\I}{\pi}\ln\frac{(\omega-\omega_{max})^2+\omega_T^2}{(\omega+\omega_{max})^2}\right)\Big]\;,
\end{align}
\end{widetext}
where $\omega_\mu=2\mu/\hbar$, $\omega_\Delta=\gamma\Delta/\hbar$ and $\omega_{max}=	\max(\omega_\mu,\omega_\Delta)$, as defined above.  Further, we introduced the damping rate $\gamma_\tau=1/\tau$ with $\tau$ the electronic relaxation time.

\subsection{Universal absorption}
\label{UniversalAbsorption}
For a two-dimensional layer, the absorption can be defined as 
\begin{equation}
\mathcal{A}=\frac{W_a}{W_i}\;,
\end{equation}
where $W_i,W_a$ are the incoming and absorbed energy flux, respectively. The absorbed energy flux is related to the average power dissipation in the layer which is proportional to the product of the local electric field at the graphene layer and the induced current. In Fourier space, these two quantities are related by the conductivity via $\j(\omega)=\sigma(\omega)\E(\omega)$, and we have $W_a=\re\sigma(\omega)|E(\omega)|^2$.
 
For incident light in air, the local electric field amplitude at the two-dimensional interface is given by $|\E|=(1+r)|\E_0|$ where $\E_0$ the incident field and $r$ the Fresnel reflection coefficient at the interface. With the incoming flux of a linearly polarized wave $W_i=\frac{\epsilon_0c}{2}|\E_0|^2$, the general graphene absorption can then be written as
\begin{align}
\mathcal{A}=|1+r|^2\frac{\re \sigma}{\epsilon_0c}\;.
\end{align}
This formula holds for an arbitrary multilayer substrate. For a single interface with $t=1+r$ the Fresnel transmission coefficient, we have $t=\frac{2}{1+n_s}$ with $n_s$ the refractive index of the substrate.

For transitions at the band gap at $T=0$ and $\gamma_\tau=0$, we thus obtain the universal absorption to be
\begin{align}
\label{UniversalAbs}
\mathcal{A}=g_sg_v\nu|1+r|^2\mathcal{A}_0
\end{align}
with the absorption quantum $\mathcal{A}_0=\pi\alpha/4$ and $\alpha\equiv1/137$ the fine-structure constant. With $\nu$, we again denote the dispersion close to the band edge  which is $\nu=2n$ for $\Delta\neq0$ and and $\nu=n$ for $\Delta=0$. 

With $\nu=1$, $g_s=2$ and $g_v=2$, we obtain the well-known absorption of $\mathcal{A}=\pi\alpha$ for suspended graphene, whereas with $\nu=2$, $g_s=2$ and $g_v=1$ and $(1+r)=t=\frac{2}{1+n_s}$ with $n_s$ the refractive index of the substrate, we obtain the final result of Ref. \onlinecite{Fang13}, i.e., the absorption of a InAs-monolayer on top of a dielectric.

Eq. (\ref{UniversalAbs}) represents the basic result of this work. To demonstrate that the same result is also obtained for Schr\"odinger electrons, we will calculate the optical conductivity based on a general $\k\cdot\p$-model in the next section.

\iffalse
\begin{figure}
\centering
  \includegraphics[width=0.99\columnwidth]{fig4c.pdf}
\caption{(color online): Comparison of the experimental data of Ref. \onlinecite{DiPietro13} (symbols) with the optical mode of Eq. (\ref{OpticalModeTwo}). Right: Including only Dirac Fermions with $\zeta=0$, $d=60$nm (black), $d=120$nm (red) and $\zeta=0.9$, $d=120$nm  (blue). Also shown the optical mode of Eq. (\ref{OpticalMode}) (dashed). Right: Including Dirac Fermions and 2DEG with $\zeta=0$, $d=60$nm (black), $d=120$nm (red) for $\epsilon_B=10$ (full lines) and $\epsilon_B=6$ (dashed lines).}
  \label{Compare}
\end{figure}
\fi 

\section{Absorption in a semiconductor}

In this section, we will consider a general semiconductor with $H_0=\frac{\p^2}{2m_0}+V(\R)$ where $\p=-i\hbar\partial_\R$ and the periodic potential $V(\R)=V(\R+\a_i)$ along the lattice constants $\a_i$. 

\subsection{Basic model}
From Bloch's theorem, we can write the wave function as $\psi_\k(\R)=e^{i\k\cdot\R}u_{n\k}(\R)$ with $\k$ denoting the Bloch wave vector. The effective Hamiltonian for the periodic function $u_{n\k}(\R)=u_{n\k}(\R+\a_i)$ is thus given by
\begin{align}
H_{\k\cdot\p}(\k)=H_0+\frac{\hbar}{m_0}\k\cdot\p+\frac{\hbar^2k^2}{2m_0}\;.
\end{align}
This Hamiltonian shall be represented within a minimal basis set consisting of $|s\rangle$ for the conduction band and $|p_i\rangle$ for the valence band with $i=x,y,z$ which correspond to the Bloch function $u_{n\k}$ at $\k=0$. To model dichalcogenides, the relevant bands would be mainly formed by $d$-oribtals with a small influence of $p$-orbitals.\cite{Xiao12,Cappelluti13,Kormanyos13} With $\epsilon_0(k)=\frac{\hbar^2k^2}{2m_0}$ and $\langle s|\p|p_i\rangle\equiv i\frac{m_0}{\hbar}P$, we can thus write $H_{\k\cdot\p}(\k)=$
\begin{align}
\left(
\begin{array}{cccc}
\epsilon_c+\epsilon_0(k)&iPk_x&iPk_y&iPk_z\\
-iPk_x&\epsilon_v+\epsilon_0(k)&0&0\\
-iPk_y&0&\epsilon_v+\epsilon_0(k)&0\\
-iPk_z&0&0&\epsilon_v+\epsilon_0(k)
\end{array}
\right)\;.
\end{align}
The valence band splits into a light hole with energy $\epsilon_{lh}(k)=\frac{1}{2}(\epsilon_c+\epsilon_v)+\epsilon_0(k)-\sqrt{E_g^2/4+P^2k^2}$ and a doubly degenerated heavy hole with energy $\epsilon_{hh}=\epsilon_v+\epsilon_0(k)$ where $E_g=\epsilon_c-\epsilon_v$.
The energy of the conduction band is renormalized to $\epsilon_e(k)=\frac{1}{2}(\epsilon_c+\epsilon_v)+\epsilon_0(k)+\sqrt{E_g^2/4+P^2k^2}$.

Let us neglect the degenerate heavy hole band and approximate the other two bands for small $\k$. This yields 
\begin{align}
\epsilon_e(k)=\epsilon_c+\frac{\hbar^2k^2}{2m_0m_e}\;,\;\epsilon_{lh}(k)=\epsilon_v-\frac{\hbar^2k^2}{2m_0m_{lh}}\;,
\end{align}
with the effective (dimensionless) masses $m_e^{-1}=E_P/E_g+1$ and $m_{lh}^{-1}=E_P/E_g-1$ where $E_P=2m_0P^2/\hbar^2$. The reduced mass is thus given by $m_e^{-1}+m_{lh}^{-1}=2E_P/E_g$ which is the crucial relation in order to obtain a universal optical conductivity for 2D semiconductor. 

\subsection{Transition matrix element}
For the optical conductivity or absorption, we need to evaluate the transition matrix element $\langle c\k|\e_0\cdot\p|v\k'\rangle$ where we only consider transitions from the valence ($v$) to the conduction ($c$) band. $\e_0$ denotes the direction of the linearly polarized incident light. Using the above model, the full wave function is the product of the envelope function with the Bloch function at $\k=0$, $\psi_{\k}(\R)\propto\chi_\k(\R)u_{\k=0}(\R)$. The envelope function varies over a much longer scale than the unit cell and we can approximately write
\begin{align}
\langle \k c|\e_0\cdot\p|\k' v\rangle 
&\approx \e_0\cdot\p_{c,v}(0)\int\chi_{c\k}^*(\R)\chi_{v\k'}(\R)d^3R
\end{align} 
For a quantum well, the envelope function can be written as $\chi_n(\r)=A^{-1/2}e^{i\k\cdot\r}\phi_{n}(z)$. For these systems the matrix element reads
\begin{align}
&\langle \k nc|\e_0\cdot\p|\k' mv\rangle\approx 
\e_0\cdot\p_{nc,mv}(0)\delta_{\k,\k'}\\\notag
&\times\int\phi_{nc}^*(z)\phi_{mv}(z)dz\equiv p_{nc,mv}\delta_{\k,\k'}\langle nc|mv\rangle\;.
\end{align}

\subsection{Conductivity}
We can now discuss the real part of the conductivity of a 3D semiconductor which is given by
\begin{align}
\re\sigma_{reg}(\omega)&\approx\frac{\pi e^2}{m_0^2\omega}|p_{cv}|^2\frac{g_sg_v}{V}\sum_\k\notag\\
&\times\delta(\epsilon_c(\k)-\epsilon_v(\k)-\hbar\omega)\;.
\end{align}
Inserting the specific envelope function, the 2D version then reads
\begin{align}
\re\sigma_{reg}(\omega)&\approx\frac{\pi e^2}{m_0^2L\omega}|p_{cn,vm}|^2|\langle cn|vm\rangle|^2\\\notag
&\times\frac{g_sg_v}{A}\sum_\k\delta(\epsilon_{c,n}(\k)-\epsilon_{v,m}(\k)-\hbar\omega)
\end{align}
where the energy bands for small $\k$ can be approximated by $\epsilon_{b,n}(
k)=\epsilon_b\pm\epsilon_{b,n}\pm\frac{\hbar^2k^2}{2m_0m_{b,n}}$ and the upper and lower sign stands for the conduction ($b=c$) and valence ($b=v$), respectively. 
With the joint density-of-states
\begin{align}
\frac{g_sg_v}{A}\sum_\k\delta(\epsilon_{c,n}(\k)-\epsilon_{v,m}(\k)-\hbar\omega)=\\\notag
\frac{g_sg_vm_0m_{nm}}{\pi\hbar^2}\theta\left[\hbar\omega-(E_g+\epsilon_{c,n}+\epsilon_{v,n})\right]
\end{align}
where $m_{nm}^{-1}=m_{c,n}^{-1}+m_{v,m}^{-1}$ and $E_g=\epsilon_c-\epsilon_v$ the energy gap, the absorption shows a step-like behavior as function of the photon energy as more and more transitions from different sub-bands are involved. The height of these steps is quasi-universal if we assume  $p_{cn,vm}=im_0P/\hbar$, $\langle cn|vm\rangle\approx1$ and $\hbar\omega\approx E_g$:
\begin{align}
\Delta\re\sigma=\frac{g_sg_ve^2}{8\hbar L}\equiv\frac{g_sg_v2\sigma_0}{L}
\end{align}
Apart from the geometrical factor $L$, this is the same result as for graphene, and has already been noted and discussed in Ref. \onlinecite{Davies05}. But note that only transitions at the $\Gamma$-point, i.e., one valley with $g_v=1$, are involved.  We have twice the absorption of graphene per channel, consistent with the fact, the the curvature around the band gap is given by $\nu=2$. 

\subsection{Fermi's Golden rule and absorption}
We can also discuss the absorption using Fermi's Golden rule which is an alternative way to the procedure outlined in Sec. \ref{UniversalAbsorption}. For the Hamiltonian $H_0=\frac{\p^2}{2m_0}+V(\R)$, the Peierls substitution $\p\to\p+e\A(t)$ leads to a paramagnetic  as well as to a diamagnetic perturbation, $H_{par}=\frac{\A\cdot\p}{m_0}$ and $H_{dia}= \frac{\A^2}{2m_0}$, respectively. The contribution of the diamagnetic term does not contribute at finite frequencies.  Parameterizing the gauge potential as $\E(t)=-\partial_t\A=\E_0\cos(\omega t)$, the time-dependent perturbation thus reads $V(t)=\frac{eE_0}{im_0\omega}{\bm p}\cdot{\bm e}_0\sin(\omega t)$ where we defined $\E_0={\bm e}_0E_0$. If we only consider transitions from the valence band $\epsilon_v(\k)$ to the conduction band $\epsilon_c(\k)$, the absorbed energy obtained from Fermi's Golden rule is obtained as
\begin{align}
\frac{W_a}{E_0^2}&=\frac{\pi}{2}\frac{e^2}{m_0^2\omega}\frac{g_sg_v}{V}\sum_\k\left|\langle \k,c|{\bm e}_0\cdot{\bm p}|\k,v\rangle\right|^2\\\notag
&\times\delta(\epsilon_c(\k)-\epsilon_v(\k)-\hbar\omega)\;.
\end{align}
We thus obtain the following absorption for the (suspended) system:
\begin{align}
\mathcal{A}=g_sg_v2\mathcal{A}_0
\end{align}
Again, we read off the band-curvature to be $\nu=2$ and with $g_s=2$ and $g_v=1$, this agrees with the absorption $\mathcal{A}=\pi\alpha$ of suspended graphene where $g_s=g_v=2$ and $\nu=1$.

\section{Hybrid model of Dirac and Schr\"odinger fermions}
\label{FullModel}
Let us finally investigate a model Hamiltonian of chiral massive electrons with a $k$-dependent mass term:
\begin{equation}
\label{HybridModel}
H= E_0+\alpha k^l+\gamma  \left( \begin{array}{cc}
\Delta/2 +\beta k^m& (k_x -  i k_y)^n  \\
(k_x +i k_y)^n  & -\Delta/2-\beta k^m 
\end{array} \right) \;.
\end{equation}
To be more general, we also included a constant shift $E_0$ as well as an isotropic $k$-dependent potential. The eigenvalues are given by $\epsilon_\k=E_0+\alpha k^l\pm\gamma\sqrt{k^{2n}+(\Delta/2+\beta k^m)^2}$.
The spinor eigenvectors are again given by Eq. (\ref{eigenvectors}), but this time 
 with $\cos\vartheta=\frac{\Delta+2\beta k^m}{\sqrt{4k^{2n}+ (\Delta+2\beta k^m)^2}}$. The velocity matrix now also has non-diagonal entries. After integration over the angle, one obtains
\begin{align}
\frac{1}{2\pi}&\int_0^{2\pi}|\langle \k,+|v^i|\k,-\rangle|^2=\frac{1}{2}\left(\frac{n\gamma k^{n-1}}{\hbar}\right)^2\\\notag&\times\left(1+\left(\frac{m}{n}\beta k^{m-n}\sin\vartheta-\cos\vartheta\right)^2\right)\;.
\end{align}

\subsection{Dissipative response}
Let us discuss the real part of the conductivity due to intra- and interband contribution and consider two special cases where $m=n$ and $m=2n$. Again, we do not have to make explicit reference to the diamagnetic current by using Eq. (\ref{rhorho}) and the continuity equation, valid only for the real - paramagnetic plus diamagnetic - current. Still, in the subsequent subsection, we will also discuss the diamagnetic current for the general model.

Regarding the Drude weight, $D$, we will only present results for the special case $E_0=\alpha=0$ even though for $l=m$, the calculation of $D$ is straightforward for the two special cases. But the full expressions are lengthy and one does not gain much insight. The results for $\re\sigma_{reg}$, though, hold for the general model with arbitrary $l$ and we will express the results with respect to the dimensionless frequency $\Omega=\omega/\omega_\Delta$.

For $m=n$, we have the following Drude weight per channel ($E_0=\alpha=0$):
\begin{align}
\frac{D}{2m\sigma_0\omega_\mu}&=1-\frac{\omega_\Delta^2}{(1+\beta^2)\omega_\mu^2}\\\notag
&\times\Big(1+\beta\sqrt{(\omega_\mu/\omega_\Delta)^2-1+\beta^2}\Big)\;.
\end{align}
For the real part of the regular conductivity, we obtain the following result:
\begin{align}
\frac{\re\sigma_{reg}}{m\sigma_0}=\frac{(\sqrt{\Omega^2-1+\beta^2}-\beta)(1+\Omega^2)}{(1+\beta^2)\sqrt{\Omega^2-1+\beta^2}}\theta(\omega-\omega_{max})\;,
\end{align}
with $\omega_{max}=\max(\omega_{\tilde\mu},\omega_\Delta)$ and $\omega_{\tilde\mu}=2\tilde\mu/\hbar$ where we introduced the shifted chemical potential $\tilde\mu=|\mu-E_0-\alpha k_F^l|$ and $k_F$ denotes the Fermi wave vector ($k_F=0$ for half-filling). Interestingly, there is no optical conductivity and thus no optical absorption for transitions at the band-edge and half-filling. 

For $m=2n$, we have the following Drude weight per channel ($E_0=\alpha=0$):
\begin{align}
\frac{D}{2m\sigma_0\omega_\mu}= &1+\frac{\omega_\beta^2}{\omega_\mu^2}\Big(1+2\Delta\beta-(1+\Delta\beta)\\\notag
&\times\sqrt{1+2\Delta\beta+(\omega_\mu/\omega_\beta)^2}\Big)\;,
\end{align}
where we defined $\omega_\beta=\gamma/\beta/\hbar$. Including temperature, we obtain the same expression as for $\beta=0$ to first order in $k_BT$, given in Eq. (\ref{FiniteTemp}).

For the real part of the regular conductivity, we obtain the following result first derived in Refs. \onlinecite{Peres13,Juergens14,Rostami14} for $n=1$:
\begin{align}
\frac{\re\sigma_{reg}}{m\sigma_0}&=\frac{\theta(\omega-\omega_{max})}{\sqrt{1+2\Delta\beta+\Omega^2}}\Big(1+\frac{1+2\Delta\beta}{\Omega^2}\Big[1+\Delta\beta\\\notag
&-\sqrt{1+2\Delta\beta+\Omega^2}\Big]\Big)
\end{align}
At the band-edge and for half-filling, this becomes
\begin{align}
\re\sigma_{reg}(\omega=\omega_\Delta)=\frac{2n\sigma_0}{1+\Delta\beta}\;.
\end{align}
There is thus a non-universal absorption depending on the product of the band-gap $\Delta$ and the mixing parameter $\beta$. For parameters of MoS$_2$,\cite{Rostami13,Kormanyos13,Scholz13} we obtain $\Delta\beta\approx0.84$ and thus $\sigma\approx\sigma_0$. There is thus a clear difference modelling MoS$_2$ with or without the mixing parameter $\beta$ in the optical bulk absorption. We note, though, that for the true absorption of MoS$_2$,\cite{Mak10} excitonic effects are important which are neglected in this one-particle approach.\cite{Rama12,Qiu13}

For parameters of Te(Cd)Hg-quantum wells, we obtain $\Delta\beta\approx0.04$ and there is thus only a negligible effect of the mixing parameter on the universal absorption. Still, we see that the optical conductivity is enhanced in the topologically non-trivial phase $\Delta<0$ in which the optical conductivity  even diverges for $\Delta\beta\to-1$.\cite{Peres13,Juergens14,Rostami14}

\subsection{Diamagnetic current}
\label{FullModelDia}
In order to complete the discussion, we will also calculate the diamagnetic current for the general model of Eq. (\ref{HybridModel}). It is given by
\begin{align}
\J_{dia}(\r)=-e\langle\psi^\dag(\r)\v_{dia}\psi(\r)\rangle\;,
\end{align}
where the field operator $\psi(\r)$ is defined as usual and the diamagnetic velocity operator is obtained from Eq. (\ref{velocity}) via the Peierls substitution $\k\rightarrow\k+\frac{e}{\hbar} \A$ as the operator linear in the gauge field, $\A$.

When averaging over the ground-state, no contribution from the non-diagonal chiral part arises due to angular integration of the integrant $e^{\pm\I n\phi}$. Only Schr\"odinger electrons thus contribute and one finds
\begin{align}
\J_{dia}&=-\frac{e^2\A}{4\pi\hbar^2}\Big[\int_0^{k_F}dk\left(\alpha l^2k^{l-1}+\gamma\beta m^2k^{m-1}\cos\vartheta\right)\notag\\
&+\int_0^\Lambda dk\left(\alpha l^2k^{l-1}-\gamma\beta m^2k^{m-1}\cos\vartheta\right)\Big],
\end{align}
where $k_F$ is the Fermi wave vector in the conduction band 
corresponding to the chemical potential $\mu$, and $\Lambda$ is a wave vector cutoff in the valence band which can be related to the carrier density of the undoped system.

Let us again emphazise that there is no diamagnetic current for pure chiral fermions independent of $n$ (for $n=1$, this statement would be, of course, trivial). This peculiarity does not lead to a violation of the $f$-sum rule which in tight-binding models connects the spectral weight to the diamagnetic term and is a consequence of charge conservation.\cite{Gusynin07,Benfatto08,Stauber10} With
respect to continuous models, this has been discussed for single layer\cite{Sabio08} as well as of twisted bilayer\cite{Stauber13} graphene and yields a spectral weight proportional to the band cutoff $\Lambda$. The sum rule of the continuous hybrid model of Dirac and Schr\"odinger electrons shows a logarithmic dependence on the band cutoff, $\ln\Lambda$.\cite{JuergensPRB14}

\section{Conclusions}
We have discussed the optical response of general two-band models and have argued that a universal optical conductivity can be defined for two-dimensional systems which are composed of pure Dirac (chiral) or Schr\"odinger electrons. For hybrid systems, present in HgTe/(Hg,Cd)Te-quantum wells or MoS$_2$, we find non-universal behaviour.

Our results point at an intriguing interplay of light-matter interaction which deserves further attention. Since the fundamental light-matter coupling
is defined by the fine-structure constant $\alpha$, one would naturally expect the absorption of 2D systems to be proportional to this constant since the scattering rate is governed by $\alpha$. Still, an open question remains why the prefactor $\pi/4$ appears in the absorption quantum $\mathcal{A}_0$ and whether it is related to some more fundamental (geometrical) relation\cite{Ludwig94} or even to the correction of the $g$-factor which is $\alpha/2\pi$.\cite{Schwinger48}

\begin{acknowledgments} 
We thank Guillermo G\'omez Santos for useful discussions. This work has been supported by MINECO under grant FIS2013-48048-P, and by Deutsche
Forschungsgemeinschaft via GRK 1570.
\end{acknowledgments} 
\bibliography{grapheneOptics} % Produces the bibliography via BibTeX.

\begin{thebibliography}{40}%
\makeatletter
\providecommand \@ifxundefined [1]{%
 \@ifx{#1\undefined}
}%
\providecommand \@ifnum [1]{%
 \ifnum #1\expandafter \@firstoftwo
 \else \expandafter \@secondoftwo
 \fi
}%
\providecommand \@ifx [1]{%
 \ifx #1\expandafter \@firstoftwo
 \else \expandafter \@secondoftwo
 \fi
}%
\providecommand \natexlab [1]{#1}%
\providecommand \enquote  [1]{``#1''}%
\providecommand \bibnamefont  [1]{#1}%
\providecommand \bibfnamefont [1]{#1}%
\providecommand \citenamefont [1]{#1}%
\providecommand \href@noop [0]{\@secondoftwo}%
\providecommand \href [0]{\begingroup \@sanitize@url \@href}%
\providecommand \@href[1]{\@@startlink{#1}\@@href}%
\providecommand \@@href[1]{\endgroup#1\@@endlink}%
\providecommand \@sanitize@url [0]{\catcode `\\12\catcode `\$12\catcode
  `\&12\catcode `\#12\catcode `\^12\catcode `\_12\catcode `\%12\relax}%
\providecommand \@@startlink[1]{}%
\providecommand \@@endlink[0]{}%
\providecommand \url  [0]{\begingroup\@sanitize@url \@url }%
\providecommand \@url [1]{\endgroup\@href {#1}{\urlprefix }}%
\providecommand \urlprefix  [0]{URL }%
\providecommand \Eprint [0]{\href }%
\providecommand \doibase [0]{http://dx.doi.org/}%
\providecommand \selectlanguage [0]{\@gobble}%
\providecommand \bibinfo  [0]{\@secondoftwo}%
\providecommand \bibfield  [0]{\@secondoftwo}%
\providecommand \translation [1]{[#1]}%
\providecommand \BibitemOpen [0]{}%
\providecommand \bibitemStop [0]{}%
\providecommand \bibitemNoStop [0]{.\EOS\space}%
\providecommand \EOS [0]{\spacefactor3000\relax}%
\providecommand \BibitemShut  [1]{\csname bibitem#1\endcsname}%
\let\auto@bib@innerbib\@empty
%</preamble>
\bibitem [{\citenamefont {Mak}\ \emph {et~al.}(2008)\citenamefont {Mak},
  \citenamefont {Sfeir}, \citenamefont {Wu}, \citenamefont {Lui}, \citenamefont
  {Misewich},\ and\ \citenamefont {Heinz}}]{Mak08}%
  \BibitemOpen
  \bibfield  {author} {\bibinfo {author} {\bibfnamefont {K.~F.}\ \bibnamefont
  {Mak}}, \bibinfo {author} {\bibfnamefont {M.~Y.}\ \bibnamefont {Sfeir}},
  \bibinfo {author} {\bibfnamefont {Y.}~\bibnamefont {Wu}}, \bibinfo {author}
  {\bibfnamefont {C.~H.}\ \bibnamefont {Lui}}, \bibinfo {author} {\bibfnamefont
  {J.~A.}\ \bibnamefont {Misewich}}, \ and\ \bibinfo {author} {\bibfnamefont
  {T.~F.}\ \bibnamefont {Heinz}},\ }\href@noop {} {\bibfield  {journal}
  {\bibinfo  {journal} {Phys. Rev. Lett.}\ }\textbf {\bibinfo {volume} {101}},\
  \bibinfo {pages} {196405} (\bibinfo {year} {2008})}\BibitemShut {NoStop}%
\bibitem [{\citenamefont {Nair}\ \emph {et~al.}(2008)\citenamefont {Nair},
  \citenamefont {Blake}, \citenamefont {Grigorenko}, \citenamefont {Novoselov},
  \citenamefont {Booth}, \citenamefont {Stauber}, \citenamefont {Peres},\ and\
  \citenamefont {Geim}}]{Nair08}%
  \BibitemOpen
  \bibfield  {author} {\bibinfo {author} {\bibfnamefont {R.~R.}\ \bibnamefont
  {Nair}}, \bibinfo {author} {\bibfnamefont {P.}~\bibnamefont {Blake}},
  \bibinfo {author} {\bibfnamefont {A.~N.}\ \bibnamefont {Grigorenko}},
  \bibinfo {author} {\bibfnamefont {K.~S.}\ \bibnamefont {Novoselov}}, \bibinfo
  {author} {\bibfnamefont {T.~J.}\ \bibnamefont {Booth}}, \bibinfo {author}
  {\bibfnamefont {T.}~\bibnamefont {Stauber}}, \bibinfo {author} {\bibfnamefont
  {N.~M.~R.}\ \bibnamefont {Peres}}, \ and\ \bibinfo {author} {\bibfnamefont
  {A.~K.}\ \bibnamefont {Geim}},\ }\href@noop {} {\bibfield  {journal}
  {\bibinfo  {journal} {Science}\ }\textbf {\bibinfo {volume} {320}},\ \bibinfo
  {pages} {1308} (\bibinfo {year} {2008})}\BibitemShut {NoStop}%
\bibitem [{\citenamefont {Ludwig}\ \emph {et~al.}(1994)\citenamefont {Ludwig},
  \citenamefont {Fisher}, \citenamefont {Shankar},\ and\ \citenamefont
  {Grinstein}}]{Ludwig94}%
  \BibitemOpen
  \bibfield  {author} {\bibinfo {author} {\bibfnamefont {A.~W.~W.}\
  \bibnamefont {Ludwig}}, \bibinfo {author} {\bibfnamefont {M.~P.~A.}\
  \bibnamefont {Fisher}}, \bibinfo {author} {\bibfnamefont {R.}~\bibnamefont
  {Shankar}}, \ and\ \bibinfo {author} {\bibfnamefont {G.}~\bibnamefont
  {Grinstein}},\ }\href {\doibase 10.1103/PhysRevB.50.7526} {\bibfield
  {journal} {\bibinfo  {journal} {Phys. Rev. B}\ }\textbf {\bibinfo {volume}
  {50}},\ \bibinfo {pages} {7526} (\bibinfo {year} {1994})}\BibitemShut
  {NoStop}%
\bibitem [{\citenamefont {Ando}\ \emph {et~al.}(2002)\citenamefont {Ando},
  \citenamefont {Zheng},\ and\ \citenamefont {Suzuura}}]{Ando02}%
  \BibitemOpen
  \bibfield  {author} {\bibinfo {author} {\bibfnamefont {T.}~\bibnamefont
  {Ando}}, \bibinfo {author} {\bibfnamefont {Y.}~\bibnamefont {Zheng}}, \ and\
  \bibinfo {author} {\bibfnamefont {H.}~\bibnamefont {Suzuura}},\ }\href
  {\doibase 10.1143/JPSJ.71.1318} {\bibfield  {journal} {\bibinfo  {journal}
  {Journal of the Physical Society of Japan}\ }\textbf {\bibinfo {volume}
  {71}},\ \bibinfo {pages} {1318} (\bibinfo {year} {2002})}\BibitemShut
  {NoStop}%
\bibitem [{\citenamefont {Gusynin}\ and\ \citenamefont
  {Sharapov}(2006)}]{Gusynin06}%
  \BibitemOpen
  \bibfield  {author} {\bibinfo {author} {\bibfnamefont {V.~P.}\ \bibnamefont
  {Gusynin}}\ and\ \bibinfo {author} {\bibfnamefont {S.~G.}\ \bibnamefont
  {Sharapov}},\ }\href {\doibase 10.1103/PhysRevB.73.245411} {\bibfield
  {journal} {\bibinfo  {journal} {Phys. Rev. B}\ }\textbf {\bibinfo {volume}
  {73}},\ \bibinfo {pages} {245411} (\bibinfo {year} {2006})}\BibitemShut
  {NoStop}%
\bibitem [{\citenamefont {Falkovsky}\ and\ \citenamefont
  {Pershoguba}(2007)}]{FalkovskyPer07}%
  \BibitemOpen
  \bibfield  {author} {\bibinfo {author} {\bibfnamefont {L.~A.}\ \bibnamefont
  {Falkovsky}}\ and\ \bibinfo {author} {\bibfnamefont {S.~S.}\ \bibnamefont
  {Pershoguba}},\ }\href {\doibase 10.1103/PhysRevB.76.153410} {\bibfield
  {journal} {\bibinfo  {journal} {Phys. Rev. B}\ }\textbf {\bibinfo {volume}
  {76}},\ \bibinfo {pages} {153410} (\bibinfo {year} {2007})}\BibitemShut
  {NoStop}%
\bibitem [{\citenamefont {Peres}\ and\ \citenamefont
  {Stauber}(2008)}]{Peres08}%
  \BibitemOpen
  \bibfield  {author} {\bibinfo {author} {\bibfnamefont {N.~M.~R.}\
  \bibnamefont {Peres}}\ and\ \bibinfo {author} {\bibfnamefont
  {T.}~\bibnamefont {Stauber}},\ }\href {\doibase 10.1142/S0217979208039794}
  {\bibfield  {journal} {\bibinfo  {journal} {Int. J. Mod. Phys. B}\ }\textbf
  {\bibinfo {volume} {22}},\ \bibinfo {pages} {2529} (\bibinfo {year}
  {2008})}\BibitemShut {NoStop}%
\bibitem [{\citenamefont {Stauber}\ \emph {et~al.}(2008)\citenamefont
  {Stauber}, \citenamefont {Peres},\ and\ \citenamefont {Geim}}]{Stauber08}%
  \BibitemOpen
  \bibfield  {author} {\bibinfo {author} {\bibfnamefont {T.}~\bibnamefont
  {Stauber}}, \bibinfo {author} {\bibfnamefont {N.~M.~R.}\ \bibnamefont
  {Peres}}, \ and\ \bibinfo {author} {\bibfnamefont {A.~K.}\ \bibnamefont
  {Geim}},\ }\href {\doibase 10.1103/PhysRevB.78.085432} {\bibfield  {journal}
  {\bibinfo  {journal} {Phys. Rev. B}\ }\textbf {\bibinfo {volume} {78}},\
  \bibinfo {pages} {085432} (\bibinfo {year} {2008})}\BibitemShut {NoStop}%
\bibitem [{\citenamefont {Mishchenko}(2008)}]{Mishchenko08}%
  \BibitemOpen
  \bibfield  {author} {\bibinfo {author} {\bibfnamefont {E.~G.}\ \bibnamefont
  {Mishchenko}},\ }\href {http://stacks.iop.org/0295-5075/83/i=1/a=17005}
  {\bibfield  {journal} {\bibinfo  {journal} {EPL (Europhysics Letters)}\
  }\textbf {\bibinfo {volume} {83}},\ \bibinfo {pages} {17005} (\bibinfo {year}
  {2008})}\BibitemShut {NoStop}%
\bibitem [{\citenamefont {Sheehy}\ and\ \citenamefont
  {Schmalian}(2009)}]{Sheehy09}%
  \BibitemOpen
  \bibfield  {author} {\bibinfo {author} {\bibfnamefont {D.~E.}\ \bibnamefont
  {Sheehy}}\ and\ \bibinfo {author} {\bibfnamefont {J.}~\bibnamefont
  {Schmalian}},\ }\href {\doibase 10.1103/PhysRevB.80.193411} {\bibfield
  {journal} {\bibinfo  {journal} {Phys. Rev. B}\ }\textbf {\bibinfo {volume}
  {80}},\ \bibinfo {pages} {193411} (\bibinfo {year} {2009})}\BibitemShut
  {NoStop}%
\bibitem [{\citenamefont {Teber}\ and\ \citenamefont
  {Kotikov}(2014)}]{Teber14}%
  \BibitemOpen
  \bibfield  {author} {\bibinfo {author} {\bibfnamefont {S.}~\bibnamefont
  {Teber}}\ and\ \bibinfo {author} {\bibfnamefont {A.~V.}\ \bibnamefont
  {Kotikov}},\ }\href {http://stacks.iop.org/0295-5075/107/i=5/a=57001}
  {\bibfield  {journal} {\bibinfo  {journal} {EPL (Europhysics Letters)}\
  }\textbf {\bibinfo {volume} {107}},\ \bibinfo {pages} {57001} (\bibinfo
  {year} {2014})}\BibitemShut {NoStop}%
\bibitem [{\citenamefont {Fang}\ \emph {et~al.}(2013)\citenamefont {Fang},
  \citenamefont {Bechtel}, \citenamefont {Plis}, \citenamefont {Martin},
  \citenamefont {Krishna}, \citenamefont {Yablonovitch},\ and\ \citenamefont
  {Javey}}]{Fang13}%
  \BibitemOpen
  \bibfield  {author} {\bibinfo {author} {\bibfnamefont {H.}~\bibnamefont
  {Fang}}, \bibinfo {author} {\bibfnamefont {H.~A.}\ \bibnamefont {Bechtel}},
  \bibinfo {author} {\bibfnamefont {E.}~\bibnamefont {Plis}}, \bibinfo {author}
  {\bibfnamefont {M.~C.}\ \bibnamefont {Martin}}, \bibinfo {author}
  {\bibfnamefont {S.}~\bibnamefont {Krishna}}, \bibinfo {author} {\bibfnamefont
  {E.}~\bibnamefont {Yablonovitch}}, \ and\ \bibinfo {author} {\bibfnamefont
  {A.}~\bibnamefont {Javey}},\ }\href {\doibase 10.1073/pnas.1309563110}
  {\bibfield  {journal} {\bibinfo  {journal} {Proceedings of the National
  Academy of Sciences}\ }\textbf {\bibinfo {volume} {110}},\ \bibinfo {pages}
  {11688} (\bibinfo {year} {2013})}\BibitemShut {NoStop}%
\bibitem [{\citenamefont {Hasan}\ and\ \citenamefont {Kane}(2010)}]{Hasan10}%
  \BibitemOpen
  \bibfield  {author} {\bibinfo {author} {\bibfnamefont {M.}~\bibnamefont
  {Hasan}}\ and\ \bibinfo {author} {\bibfnamefont {C.}~\bibnamefont {Kane}},\
  }\href {\doibase 10.1103/RevModPhys.82.3045} {\bibfield  {journal} {\bibinfo
  {journal} {Rev Mod Phys}\ }\textbf {\bibinfo {volume} {82}},\ \bibinfo
  {pages} {3045} (\bibinfo {year} {2010})}\BibitemShut {NoStop}%
\bibitem [{\citenamefont {Qi}\ and\ \citenamefont {Zhang}(2011)}]{Qi11}%
  \BibitemOpen
  \bibfield  {author} {\bibinfo {author} {\bibfnamefont {X.-L.}\ \bibnamefont
  {Qi}}\ and\ \bibinfo {author} {\bibfnamefont {S.-C.}\ \bibnamefont {Zhang}},\
  }\href {\doibase 10.1103/RevModPhys.83.1057} {\bibfield  {journal} {\bibinfo
  {journal} {Rev. Mod. Phys.}\ }\textbf {\bibinfo {volume} {83}},\ \bibinfo
  {pages} {1057} (\bibinfo {year} {2011})}\BibitemShut {NoStop}%
\bibitem [{\citenamefont {Rostami}\ \emph {et~al.}(2013)\citenamefont
  {Rostami}, \citenamefont {Moghaddam},\ and\ \citenamefont
  {Asgari}}]{Rostami13}%
  \BibitemOpen
  \bibfield  {author} {\bibinfo {author} {\bibfnamefont {H.}~\bibnamefont
  {Rostami}}, \bibinfo {author} {\bibfnamefont {A.~G.}\ \bibnamefont
  {Moghaddam}}, \ and\ \bibinfo {author} {\bibfnamefont {R.}~\bibnamefont
  {Asgari}},\ }\href {\doibase 10.1103/PhysRevB.88.085440} {\bibfield
  {journal} {\bibinfo  {journal} {Phys. Rev. B}\ }\textbf {\bibinfo {volume}
  {88}},\ \bibinfo {pages} {085440} (\bibinfo {year} {2013})}\BibitemShut
  {NoStop}%
\bibitem [{\citenamefont {Bergh\"auser}\ and\ \citenamefont
  {Malic}(2014)}]{Berghauser14}%
  \BibitemOpen
  \bibfield  {author} {\bibinfo {author} {\bibfnamefont {G.}~\bibnamefont
  {Bergh\"auser}}\ and\ \bibinfo {author} {\bibfnamefont {E.}~\bibnamefont
  {Malic}},\ }\href {\doibase 10.1103/PhysRevB.89.125309} {\bibfield  {journal}
  {\bibinfo  {journal} {Phys. Rev. B}\ }\textbf {\bibinfo {volume} {89}},\
  \bibinfo {pages} {125309} (\bibinfo {year} {2014})}\BibitemShut {NoStop}%
\bibitem [{\citenamefont {Fetter}\ and\ \citenamefont
  {Walecka}(2003)}]{Fetter03}%
  \BibitemOpen
  \bibfield  {author} {\bibinfo {author} {\bibfnamefont {A.~L.}\ \bibnamefont
  {Fetter}}\ and\ \bibinfo {author} {\bibfnamefont {J.~D.}\ \bibnamefont
  {Walecka}},\ }\href@noop {} {\emph {\bibinfo {title} {Quantum Theory of
  Many-Particle Systems}}}\ (\bibinfo  {publisher} {Dover Publications, New
  York},\ \bibinfo {year} {2003})\BibitemShut {NoStop}%
\bibitem [{Note1()}]{Note1}%
  \BibitemOpen
  \bibinfo {note} {We include the prefactor $\pi $ in the definition of the
  Drude weight}\BibitemShut {NoStop}%
\bibitem [{\citenamefont {McCann}\ and\ \citenamefont
  {Fal'ko}(2006)}]{McCann06}%
  \BibitemOpen
  \bibfield  {author} {\bibinfo {author} {\bibfnamefont {E.}~\bibnamefont
  {McCann}}\ and\ \bibinfo {author} {\bibfnamefont {V.~I.}\ \bibnamefont
  {Fal'ko}},\ }\href {\doibase 10.1103/PhysRevLett.96.086805} {\bibfield
  {journal} {\bibinfo  {journal} {Phys. Rev. Lett.}\ }\textbf {\bibinfo
  {volume} {96}},\ \bibinfo {pages} {086805} (\bibinfo {year}
  {2006})}\BibitemShut {NoStop}%
\bibitem [{\citenamefont {Nilsson}\ \emph {et~al.}(2006)\citenamefont
  {Nilsson}, \citenamefont {Castro~Neto}, \citenamefont {Guinea},\ and\
  \citenamefont {Peres}}]{Nilsson06}%
  \BibitemOpen
  \bibfield  {author} {\bibinfo {author} {\bibfnamefont {J.}~\bibnamefont
  {Nilsson}}, \bibinfo {author} {\bibfnamefont {A.~H.}\ \bibnamefont
  {Castro~Neto}}, \bibinfo {author} {\bibfnamefont {F.}~\bibnamefont {Guinea}},
  \ and\ \bibinfo {author} {\bibfnamefont {N.~M.~R.}\ \bibnamefont {Peres}},\
  }\href {\doibase 10.1103/PhysRevLett.97.266801} {\bibfield  {journal}
  {\bibinfo  {journal} {Phys. Rev. Lett.}\ }\textbf {\bibinfo {volume} {97}},\
  \bibinfo {pages} {266801} (\bibinfo {year} {2006})}\BibitemShut {NoStop}%
\bibitem [{\citenamefont {Koshino}\ and\ \citenamefont
  {McCann}(2009)}]{Koshino09}%
  \BibitemOpen
  \bibfield  {author} {\bibinfo {author} {\bibfnamefont {M.}~\bibnamefont
  {Koshino}}\ and\ \bibinfo {author} {\bibfnamefont {E.}~\bibnamefont
  {McCann}},\ }\href {\doibase 10.1103/PhysRevB.79.125443} {\bibfield
  {journal} {\bibinfo  {journal} {Phys. Rev. B}\ }\textbf {\bibinfo {volume}
  {79}},\ \bibinfo {pages} {125443} (\bibinfo {year} {2009})}\BibitemShut
  {NoStop}%
\bibitem [{Note2()}]{Note2}%
  \BibitemOpen
  \bibinfo {note} {This also holds true for the general model of Eq. (\ref
  {HybridModel}).}\BibitemShut {Stop}%
\bibitem [{\citenamefont {Xiao}\ \emph {et~al.}(2012)\citenamefont {Xiao},
  \citenamefont {Liu}, \citenamefont {Feng}, \citenamefont {Xu},\ and\
  \citenamefont {Yao}}]{Xiao12}%
  \BibitemOpen
  \bibfield  {author} {\bibinfo {author} {\bibfnamefont {D.}~\bibnamefont
  {Xiao}}, \bibinfo {author} {\bibfnamefont {G.-B.}\ \bibnamefont {Liu}},
  \bibinfo {author} {\bibfnamefont {W.}~\bibnamefont {Feng}}, \bibinfo {author}
  {\bibfnamefont {X.}~\bibnamefont {Xu}}, \ and\ \bibinfo {author}
  {\bibfnamefont {W.}~\bibnamefont {Yao}},\ }\href {\doibase
  10.1103/PhysRevLett.108.196802} {\bibfield  {journal} {\bibinfo  {journal}
  {Phys. Rev. Lett.}\ }\textbf {\bibinfo {volume} {108}},\ \bibinfo {pages}
  {196802} (\bibinfo {year} {2012})}\BibitemShut {NoStop}%
\bibitem [{\citenamefont {Cappelluti}\ \emph {et~al.}(2013)\citenamefont
  {Cappelluti}, \citenamefont {Rold\'an}, \citenamefont {Silva-Guill\'en},
  \citenamefont {Ordej\'on},\ and\ \citenamefont {Guinea}}]{Cappelluti13}%
  \BibitemOpen
  \bibfield  {author} {\bibinfo {author} {\bibfnamefont {E.}~\bibnamefont
  {Cappelluti}}, \bibinfo {author} {\bibfnamefont {R.}~\bibnamefont
  {Rold\'an}}, \bibinfo {author} {\bibfnamefont {J.}~\bibnamefont
  {Silva-Guill\'en}}, \bibinfo {author} {\bibfnamefont {P.}~\bibnamefont
  {Ordej\'on}}, \ and\ \bibinfo {author} {\bibfnamefont {F.}~\bibnamefont
  {Guinea}},\ }\href {\doibase 10.1103/PhysRevB.88.075409} {\bibfield
  {journal} {\bibinfo  {journal} {Phys. Rev. B}\ }\textbf {\bibinfo {volume}
  {88}},\ \bibinfo {pages} {075409} (\bibinfo {year} {2013})}\BibitemShut
  {NoStop}%
\bibitem [{\citenamefont {Korm\'anyos}\ \emph {et~al.}(2013)\citenamefont
  {Korm\'anyos}, \citenamefont {Z\'olyomi}, \citenamefont {Drummond},
  \citenamefont {Rakyta}, \citenamefont {Burkard},\ and\ \citenamefont
  {Fal'ko}}]{Kormanyos13}%
  \BibitemOpen
  \bibfield  {author} {\bibinfo {author} {\bibfnamefont {A.}~\bibnamefont
  {Korm\'anyos}}, \bibinfo {author} {\bibfnamefont {V.}~\bibnamefont
  {Z\'olyomi}}, \bibinfo {author} {\bibfnamefont {N.~D.}\ \bibnamefont
  {Drummond}}, \bibinfo {author} {\bibfnamefont {P.}~\bibnamefont {Rakyta}},
  \bibinfo {author} {\bibfnamefont {G.}~\bibnamefont {Burkard}}, \ and\
  \bibinfo {author} {\bibfnamefont {V.~I.}\ \bibnamefont {Fal'ko}},\ }\href
  {\doibase 10.1103/PhysRevB.88.045416} {\bibfield  {journal} {\bibinfo
  {journal} {Phys. Rev. B}\ }\textbf {\bibinfo {volume} {88}},\ \bibinfo
  {pages} {045416} (\bibinfo {year} {2013})}\BibitemShut {NoStop}%
\bibitem [{\citenamefont {Davies}(2005)}]{Davies05}%
  \BibitemOpen
  \bibfield  {author} {\bibinfo {author} {\bibfnamefont {J.~H.}\ \bibnamefont
  {Davies}},\ }\href@noop {} {\emph {\bibinfo {title} {The Physics of
  Low-Dimensional Semiconductors}}}\ (\bibinfo  {publisher} {Cambridge
  University Press, New York},\ \bibinfo {year} {2005})\BibitemShut {NoStop}%
\bibitem [{\citenamefont {Peres}\ and\ \citenamefont {Santos}(2013)}]{Peres13}%
  \BibitemOpen
  \bibfield  {author} {\bibinfo {author} {\bibfnamefont {N.~M.~R.}\
  \bibnamefont {Peres}}\ and\ \bibinfo {author} {\bibfnamefont {J.~E.}\
  \bibnamefont {Santos}},\ }\href
  {http://stacks.iop.org/0953-8984/25/i=30/a=305801} {\bibfield  {journal}
  {\bibinfo  {journal} {Journal of Physics: Condensed Matter}\ }\textbf
  {\bibinfo {volume} {25}},\ \bibinfo {pages} {305801} (\bibinfo {year}
  {2013})}\BibitemShut {NoStop}%
\bibitem [{\citenamefont {Juergens}\ \emph
  {et~al.}(2014{\natexlab{a}})\citenamefont {Juergens}, \citenamefont
  {Michetti},\ and\ \citenamefont {Trauzettel}}]{Juergens14}%
  \BibitemOpen
  \bibfield  {author} {\bibinfo {author} {\bibfnamefont {S.}~\bibnamefont
  {Juergens}}, \bibinfo {author} {\bibfnamefont {P.}~\bibnamefont {Michetti}},
  \ and\ \bibinfo {author} {\bibfnamefont {B.}~\bibnamefont {Trauzettel}},\
  }\href {\doibase 10.1103/PhysRevLett.112.076804} {\bibfield  {journal}
  {\bibinfo  {journal} {Phys. Rev. Lett.}\ }\textbf {\bibinfo {volume} {112}},\
  \bibinfo {pages} {076804} (\bibinfo {year} {2014}{\natexlab{a}})}\BibitemShut
  {NoStop}%
\bibitem [{\citenamefont {Rostami}\ and\ \citenamefont
  {Asgari}(2014)}]{Rostami14}%
  \BibitemOpen
  \bibfield  {author} {\bibinfo {author} {\bibfnamefont {H.}~\bibnamefont
  {Rostami}}\ and\ \bibinfo {author} {\bibfnamefont {R.}~\bibnamefont
  {Asgari}},\ }\href {\doibase 10.1103/PhysRevB.89.115413} {\bibfield
  {journal} {\bibinfo  {journal} {Phys. Rev. B}\ }\textbf {\bibinfo {volume}
  {89}},\ \bibinfo {pages} {115413} (\bibinfo {year} {2014})}\BibitemShut
  {NoStop}%
\bibitem [{\citenamefont {Scholz}\ \emph {et~al.}(2013)\citenamefont {Scholz},
  \citenamefont {Stauber},\ and\ \citenamefont {Schliemann}}]{Scholz13}%
  \BibitemOpen
  \bibfield  {author} {\bibinfo {author} {\bibfnamefont {A.}~\bibnamefont
  {Scholz}}, \bibinfo {author} {\bibfnamefont {T.}~\bibnamefont {Stauber}}, \
  and\ \bibinfo {author} {\bibfnamefont {J.}~\bibnamefont {Schliemann}},\
  }\href {\doibase 10.1103/PhysRevB.88.035135} {\bibfield  {journal} {\bibinfo
  {journal} {Phys. Rev. B}\ }\textbf {\bibinfo {volume} {88}},\ \bibinfo
  {pages} {035135} (\bibinfo {year} {2013})}\BibitemShut {NoStop}%
\bibitem [{\citenamefont {Mak}\ \emph {et~al.}(2010)\citenamefont {Mak},
  \citenamefont {Lee}, \citenamefont {Hone}, \citenamefont {Shan},\ and\
  \citenamefont {Heinz}}]{Mak10}%
  \BibitemOpen
  \bibfield  {author} {\bibinfo {author} {\bibfnamefont {K.~F.}\ \bibnamefont
  {Mak}}, \bibinfo {author} {\bibfnamefont {C.}~\bibnamefont {Lee}}, \bibinfo
  {author} {\bibfnamefont {J.}~\bibnamefont {Hone}}, \bibinfo {author}
  {\bibfnamefont {J.}~\bibnamefont {Shan}}, \ and\ \bibinfo {author}
  {\bibfnamefont {T.~F.}\ \bibnamefont {Heinz}},\ }\href {\doibase
  10.1103/PhysRevLett.105.136805} {\bibfield  {journal} {\bibinfo  {journal}
  {Phys. Rev. Lett.}\ }\textbf {\bibinfo {volume} {105}},\ \bibinfo {pages}
  {136805} (\bibinfo {year} {2010})}\BibitemShut {NoStop}%
\bibitem [{\citenamefont {Ramasubramaniam}(2012)}]{Rama12}%
  \BibitemOpen
  \bibfield  {author} {\bibinfo {author} {\bibfnamefont {A.}~\bibnamefont
  {Ramasubramaniam}},\ }\href {\doibase 10.1103/PhysRevB.86.115409} {\bibfield
  {journal} {\bibinfo  {journal} {Phys. Rev. B}\ }\textbf {\bibinfo {volume}
  {86}},\ \bibinfo {pages} {115409} (\bibinfo {year} {2012})}\BibitemShut
  {NoStop}%
\bibitem [{\citenamefont {Qiu}\ \emph {et~al.}(2013)\citenamefont {Qiu},
  \citenamefont {da~Jornada},\ and\ \citenamefont {Louie}}]{Qiu13}%
  \BibitemOpen
  \bibfield  {author} {\bibinfo {author} {\bibfnamefont {D.~Y.}\ \bibnamefont
  {Qiu}}, \bibinfo {author} {\bibfnamefont {F.~H.}\ \bibnamefont {da~Jornada}},
  \ and\ \bibinfo {author} {\bibfnamefont {S.~G.}\ \bibnamefont {Louie}},\
  }\href {\doibase 10.1103/PhysRevLett.111.216805} {\bibfield  {journal}
  {\bibinfo  {journal} {Phys. Rev. Lett.}\ }\textbf {\bibinfo {volume} {111}},\
  \bibinfo {pages} {216805} (\bibinfo {year} {2013})}\BibitemShut {NoStop}%
\bibitem [{\citenamefont {Gusynin}\ \emph {et~al.}(2007)\citenamefont
  {Gusynin}, \citenamefont {Sharapov},\ and\ \citenamefont
  {Carbotte}}]{Gusynin07}%
  \BibitemOpen
  \bibfield  {author} {\bibinfo {author} {\bibfnamefont {V.~P.}\ \bibnamefont
  {Gusynin}}, \bibinfo {author} {\bibfnamefont {S.~G.}\ \bibnamefont
  {Sharapov}}, \ and\ \bibinfo {author} {\bibfnamefont {J.~P.}\ \bibnamefont
  {Carbotte}},\ }\href {\doibase 10.1103/PhysRevB.75.165407} {\bibfield
  {journal} {\bibinfo  {journal} {Phys. Rev. B}\ }\textbf {\bibinfo {volume}
  {75}},\ \bibinfo {pages} {165407} (\bibinfo {year} {2007})}\BibitemShut
  {NoStop}%
\bibitem [{\citenamefont {Benfatto}\ \emph {et~al.}(2008)\citenamefont
  {Benfatto}, \citenamefont {Sharapov},\ and\ \citenamefont
  {Carbotte}}]{Benfatto08}%
  \BibitemOpen
  \bibfield  {author} {\bibinfo {author} {\bibfnamefont {L.}~\bibnamefont
  {Benfatto}}, \bibinfo {author} {\bibfnamefont {S.~G.}\ \bibnamefont
  {Sharapov}}, \ and\ \bibinfo {author} {\bibfnamefont {J.~P.}\ \bibnamefont
  {Carbotte}},\ }\href {\doibase 10.1103/PhysRevB.77.125422} {\bibfield
  {journal} {\bibinfo  {journal} {Phys. Rev. B}\ }\textbf {\bibinfo {volume}
  {77}},\ \bibinfo {pages} {125422} (\bibinfo {year} {2008})}\BibitemShut
  {NoStop}%
\bibitem [{\citenamefont {Stauber}\ and\ \citenamefont
  {G\'omez-Santos}(2010)}]{Stauber10}%
  \BibitemOpen
  \bibfield  {author} {\bibinfo {author} {\bibfnamefont {T.}~\bibnamefont
  {Stauber}}\ and\ \bibinfo {author} {\bibfnamefont {G.}~\bibnamefont
  {G\'omez-Santos}},\ }\href {\doibase 10.1103/PhysRevB.82.155412} {\bibfield
  {journal} {\bibinfo  {journal} {Phys. Rev. B}\ }\textbf {\bibinfo {volume}
  {82}},\ \bibinfo {pages} {155412} (\bibinfo {year} {2010})}\BibitemShut
  {NoStop}%
\bibitem [{\citenamefont {Sabio}\ \emph {et~al.}(2008)\citenamefont {Sabio},
  \citenamefont {Nilsson},\ and\ \citenamefont {Castro~Neto}}]{Sabio08}%
  \BibitemOpen
  \bibfield  {author} {\bibinfo {author} {\bibfnamefont {J.}~\bibnamefont
  {Sabio}}, \bibinfo {author} {\bibfnamefont {J.}~\bibnamefont {Nilsson}}, \
  and\ \bibinfo {author} {\bibfnamefont {A.~H.}\ \bibnamefont {Castro~Neto}},\
  }\href {\doibase 10.1103/PhysRevB.78.075410} {\bibfield  {journal} {\bibinfo
  {journal} {Phys. Rev. B}\ }\textbf {\bibinfo {volume} {78}},\ \bibinfo
  {pages} {075410} (\bibinfo {year} {2008})}\BibitemShut {NoStop}%
\bibitem [{\citenamefont {Stauber}\ \emph {et~al.}(2013)\citenamefont
  {Stauber}, \citenamefont {San-Jose},\ and\ \citenamefont {Brey}}]{Stauber13}%
  \BibitemOpen
  \bibfield  {author} {\bibinfo {author} {\bibfnamefont {T.}~\bibnamefont
  {Stauber}}, \bibinfo {author} {\bibfnamefont {P.}~\bibnamefont {San-Jose}}, \
  and\ \bibinfo {author} {\bibfnamefont {L.}~\bibnamefont {Brey}},\ }\href
  {http://stacks.iop.org/1367-2630/15/i=11/a=113050} {\bibfield  {journal}
  {\bibinfo  {journal} {New Journal of Physics}\ }\textbf {\bibinfo {volume}
  {15}},\ \bibinfo {pages} {113050} (\bibinfo {year} {2013})}\BibitemShut
  {NoStop}%
\bibitem [{\citenamefont {Juergens}\ \emph
  {et~al.}(2014{\natexlab{b}})\citenamefont {Juergens}, \citenamefont
  {Michetti},\ and\ \citenamefont {Trauzettel}}]{JuergensPRB14}%
  \BibitemOpen
  \bibfield  {author} {\bibinfo {author} {\bibfnamefont {S.}~\bibnamefont
  {Juergens}}, \bibinfo {author} {\bibfnamefont {P.}~\bibnamefont {Michetti}},
  \ and\ \bibinfo {author} {\bibfnamefont {B.}~\bibnamefont {Trauzettel}},\
  }\href {\doibase 10.1103/PhysRevB.90.115425} {\bibfield  {journal} {\bibinfo
  {journal} {Phys. Rev. B}\ }\textbf {\bibinfo {volume} {90}},\ \bibinfo
  {pages} {115425} (\bibinfo {year} {2014}{\natexlab{b}})}\BibitemShut
  {NoStop}%
\bibitem [{\citenamefont {Schwinger}(1948)}]{Schwinger48}%
  \BibitemOpen
  \bibfield  {author} {\bibinfo {author} {\bibfnamefont {J.}~\bibnamefont
  {Schwinger}},\ }\href {\doibase 10.1103/PhysRev.73.416} {\bibfield  {journal}
  {\bibinfo  {journal} {Phys. Rev.}\ }\textbf {\bibinfo {volume} {73}},\
  \bibinfo {pages} {416} (\bibinfo {year} {1948})}\BibitemShut {NoStop}%
\end{thebibliography}%
 \end{document}